# *The Einstein – Lorentz Dispute Revisited*

## *Roger Ellman*

### Abstract

Lorentz [of the Lorentz transforms and Lorentz contractions fame] contended against Einstein that there had to be a medium in which electro-magnetic waves exist and propagate, and that that would of necessity be an absolute frame of reference for the universe. Einstein won that dispute contending that electro-magnetic waves needed no medium and that there was no absolute frame of reference.

But, that victory was in a conflict of Lorentz's opinion opposed to Einstein's opinion combined with Einstein's substantial other successes and reputation. It was not a victory of solid reasoning nor demonstrated factual evidence.

Now solid reasoning and new data not available to Einstein and Lorentz show that Lorentz was correct and that Einstein's Theory of Relativity should correctly be termed Einstein's Principle of Invariance. It is shown that Einstein's comprehensive relativity and denial of an absolute frame of reference for the universe are incorrect and that the universe has an *absolute universal prime frame of reference*.

The significance of this correction in its relation to the interaction of science and society is then presented.

Roger Ellman, The-Origin Foundation, Inc.
 320 Gemma Circle, Santa Rosa, CA 95404, USA
 RogerEllman@The-Origin.org
 http://www.The-Origin.org

# The Einstein – Lorentz Dispute Revisited

## Roger Ellman

"What is motion, motion relative to what?" After all, the Earth and anything on its surface rotate about the Earth's axis, revolve around the sun, participate in the sun's motion in the galaxy and in the galaxy's motion through space. Thus use on the Earth's surface of the terms "static" or "in motion" requires clarification.

This is the fundamental problem underlying relativity, and it became a major issue upon the development of physics' treatment of electro-magnetic waves: is there a medium in which the electro-magnetic waves exist, and if so is it a "stationary" all-pervasive "aether", a prime reference system to which everything else is relative ? If not, what is the meaning of "static" or "in motion" and what of the motion of things relative to each other ?

The problem and its significance can be further appreciated by means of an example. We take a straight wire in which positive charge flows at constant velocity [constant speed and direction along the wire relative to the wire]. Classically, in terms of magnetic field behavior, there is a magnetic field circumferential to the wire. This field will exert a force on a charge moving in the field. Now, we, the observers, take on a velocity identical to the charge moving in the wire, the charge causing the magnetic field. In this case, to us, the charge in the wire is static. It is not moving and there should be no field. [It is true that to us in this case the wire appears to be traveling "rearward", but moving wires are not, in themselves, a cause of magnetic field.] Is there, now, as we view it, a magnetic field ? That is, from the "static", as we view it, charge ?

How do we reconcile this: a charge "at rest" relative to the Earth exhibits to us only static effects even though moving through space at a speed of at least 66,600 miles per hour [the Earth's speed around the sun] and a charge at rest relative to us [the above example of the wire] exhibits magnetic effects ?

### RELATIVITY AND INVARIANCE

By the time of Newton and the development of his laws of motion it was well understood that all motion is relative to some frame of reference. One cannot say that something is moving at a stated velocity except by defining what the velocity is relative to. Newtonian mechanics dealt with this problem, successfully for "Newtonian systems". Direct linear relationships transfer Newtonian motion descriptions from one frame of reference to another.

In the second half of the 19th century Maxwell developed his equations describing electro-magnetic field, the equations being an outgrowth of the then developing understanding of electricity, charge, magnetic effects, and so forth. Substantially before the first actual detection of electro-magnetic waves by Herz toward the end of the century, it was recognized that Maxwell's equations described a wave propagating in space at a velocity, $c$, determined by two constants in the equations, $\varepsilon$ and $\mu$, the dielectric constant and the permeability of whatever medium the waves were passing through, such that $c^2 = 1/\varepsilon \cdot \mu$.

This result presented two problems.

First

At the time it seemed inconceivable that these [or any] waves could propagate other than in some medium. Since the waves could and do propagate throughout free space as well as through the air and through other substances some kind of all-pervading medium, called in those days an "aether", was postulated.

Second

Maxwell's equations would not correctly transform from one frame of reference to another at different velocity using the Newtonian transformations. Therefore it was thought that Maxwell's equations applied only to one, prime, frame of reference, that of the "aether", which also defined $\mu$, $\varepsilon$, and, therefore, $c$.

[The Newtonian transform between two systems at different velocities is to merely subtract the velocity difference. For example, to a passenger in a train going forward at 30 miles per hour the train is a stationary reference system and the landscape out the window is traveling backwards at 30 miles per hour. To do a Newtonian transform from the train-as-reference to the landscape-as-reference one subtracts the landscape's 30 miles per hour backward from the landscape (making it stationary) and also from the train (making it to be going 30 miles per hour forward).

[If one attempts such a Newtonian transform on Maxwell's equations and the speed of light wrong results are obtained because of non-linearity. In addition, one cannot subtract a velocity difference between two systems from the speed of light, $c$, because $c$ is an absolute constant given by $c^2 = 1/\varepsilon \cdot \mu$ and cannot vary with some other velocity.]

The problem in the assumption that there is an "aether" which is the electro-magnetic wave medium is that all attempts to define and detect the "aether" led to contradictions or further problems. The most famous of those attempts was the Michaelson-Moreley experiment, which, expecting to find two different measured results for the speed of light because of the motion of the earth in its orbit relative to the "aether", obtained the "negative" result that the speed of light always measured to be the same regardless of the motion of the observers, Michaelson and Morely and the Earth.

The Michaelson-Moreley experiment and the Newtonian transformation inadequacy required that a new transformation system be developed. That was done by Lorentz. Lorentz retained the existence of an "aether" which had to be the prime frame of reference. His transformations and their consequent "contractions" resolved the "aether" problems. The Lorentz transforms and the Lorentz contractions are familiar to all physicists and are fundamental to the Theory of Relativity.

In the early 1900's Einstein took the further step of denying that any "aether" or medium was necessary for electro-magnetic waves and that there was no prime frame of reference. Those assumptions were embodied in his Theory of Relativity for which, there being no "aether", everything is relative. The repeated failure to successfully define and detect an "aether", coupled with Einstein's formulation that dealt with the problem by denying the "aether's" existence, resulted in the complete acceptance of Einstein's theories and the abandonment of the "aether" problem. However, Einstein had no proof, only his opinion, to justify his aether denial.

Excepting only the issue of whether an "aether" exists and is the prime frame of reference, the Lorentz and the Einstein formulations are equally valid descriptions of physical reality. However, the Theory of Relativity and other developments in physics that came from Einstein [his explanation of the photoelectric effect and his famous $E = m \cdot c^2$] were tremendously successful. Relativistic effects could be observed and measured experimentally. The mass-energy equivalence was dramatically confirmed.

Just as Einstein had his doubts about some of the then accepted aspects of traditional 20[th] Century physics [in referring to some aspects of uncertainty and quantum mechanics he is reputed to have said that he "... did not believe that God plays with dice ...."] so Lorentz still clung to the necessity of an "aether" and the prime frame of reference that it implied.

But the relativity "bandwagon" was rolling and relativity carried the day.

New developments in space research long after the death of Lorentz and Einstein now make it necessary to reverse that outcome and conclusion. It can now be shown that Lorentz was essentially correct and Einstein incorrect with regard to a prime frame of reference and a medium in which electro-magnetic waves propagate. That is, there is a universal absolute frame of reference to which all motion is relative and there is a prime frame of reference.

It is now necessary to restate relativity more correctly. There is nothing inherent in Einstein's Theory of Relativity requiring his comprehensive relativity, the absence of a prime frame of reference. The concept "relative" does not necessarily enter into the mathematical derivations and "theory of relativity" is a misnomer. The theory-system called the Theory of Relativity should be correctly referred to as the "Principle of Invariance". Einstein's postulates were solely invariance.

"Invariance" means that the laws of physics, the behavior of all physical reality, is the same in any coordinate system or frame of reference. Invariance requires that the form of the mathematical statements describing reality and the constants appearing in those statements be invariant under any transformation of coordinates, which means that they must be unchanged by any change of frame of reference regardless of its motion so long as it is at constant velocity with no acceleration involved. Since all universal constants appearing in equations describing physical reality are invariant, the speed of light, one of those constants, is invariant.

The principle of invariance is not magical or mysterious, but obvious. When one walks down the street, breathes, throws a stone or rides in a space ship one is doing a thing. The thing is not changed by changing the frame of reference from which someone observes it. The act is invariant therefore its description must be so. Einstein's principal mistake was that while he recognized that invariance was essential he did not look for a mechanism to cause that to be so, and the only possible such mechanism is a universe-wide single absolute frame of reference.

To be perfectly clear about this replacement of relativity with "absolutivity" the pertinent factors are as follows.

(a) All motion is absolute, that is, it is relative to an absolute, prime frame of reference.

> In normal human experience the absolute frame of reference cannot be detected so that motion seems to be relative, but that is only an appearance.

(b) The absolute frame of reference is not a "preferred" frame of reference in the sense of having special or different physical laws. It is a "prime" reference system in that all physical reality is relative to it.

> That is why the universe is invariant. For physical reality there is only one grand system of reference for everything. The universe does not "know" about our frames of reference; it simply is in its natural frame of reference, everywhere. It would be ridiculous for it not to be invariant.

This goes counter to some of the most basic accepted concepts of 20th Century physics. Consequently, it requires substantial justification, which is as follows.

(1) A medium is required for electro-magnetic waves. They either propagate in a medium or are themselves propagation of the wave "substance" or else they have no existence. Since they exist, and since their propagation is a transverse wave, not longitudinal, and since there has never been a contention that electro-magnetic waves involve motion of anything in the direction of wave propagation other than that of the wave's energy and momentum, the medium must exist.

One cannot say that there is no electro-magnetic wave medium just "field". "Field" is merely a code-word for "action at a distance", an inability to actually explain the mechanism and actions involved.

A medium is also required to define and set the propagation velocity of the waves to $c$, the speed of light. Without a medium there is no cause of a universal fixed value of $c$ nor $\mu$ and $\varepsilon$, the dielectric constant and permeability of free space.

(2) As described in the General Theory of Relativity, "curved" space-time, due to the variation of gravitation with the distribution of mass in the universe, and the gravitational field pervading the universe with its shape due to that variation, is itself a frame of reference. Since space-time is not uniformly "flat", the shape variations make possible detection not only of acceleration but also of absolute velocity relative to the total mass as distributed in the universe.

But, that reference frame is identical to the reference frame of the singularity at which the universe started with the "big bang".

(3) There exists throughout the universe a background radiation which is the residual radiation from the immense energy of the "big bang", the start of the universe. The temperature has now cooled down from the extremely high levels at the beginning to only about $2.7°$ $Kelvin$. That radiation is, of course, relative to the beginning, relative to "where the "big bang" took place. Measurements of Doppler frequency shift of this radiation due to the motion of the Earth give an absolute velocity for the Earth relative to the medium of about $370$ $km/sec$. The absolute direction of the Earth's motion as indicated by those measurements is off in the direction from Earth of the constellation Leo.

The absolute velocity of the Earth is sufficiently low that observations from Earth are equivalent (within the accuracy involved] to observations from at rest in the absolute frame of reference.

$$v_{Earth} \sim 370 \ ^{km}/_{sec}$$

$$\left[ 1 - \frac{v_E^2}{c^2} \right]^{½} = 0.999,999,2 \ ...$$

(4) The Lorentz contractions must actually occur, not be mere observational effects. According to the theory of relativity, an object in motion experiences slower time. If two identical clocks agree and one clock is then moved away and returned while the other is motionless [in relativistic terminology if one is moved away and then returned relative to the other from which observations are made] the moved clock must read an earlier time than the unmoved clock even when both are again at rest in the same frame of reference. When both are so again together and at rest there can be no observational quirk to cause them to read different times. The moved clock must have actually run slower.

It could be argued that the moved clock had to be accelerated to be moved so that the overall process was not a constant velocity situation. That is not the contention of relativity, however, which states that the moved clock does run slower and relies on the fact of acceleration to make the distinction as to which clock was moved and which stayed at rest.

(5) Consider three clocks, #1, #2, and #3, at constant velocities $v_1$, $v_2$, and $v_3$. According to relativity the time of clock #3 is contracted by some amount relative to Clock #1. Likewise Clock #3 is time contracted relative to Clock #2, but by some different amount. But, Clock #3, with a time contraction relative to Clock #1 in an amount based on the velocity difference between Clock #1 and Clock #3, and with a time contraction relative to Clock #2 based on the velocity difference between Clock #2 and Clock #3, cannot be actually contracted two different amounts at the same moment. Since the contraction must be actual, not solely observational, an absurdity results.

The solution to this problem is simple. All clocks are actually, as observed from the prime frame of reference, contracted according to their absolute velocity relative to that frame, not according to their velocity relative to another moving clock. In addition, an observer at a moving clock observes somewhat different results than those actual absolute contractions because his standards of measurement have also been contracted by his own motion [even though they appear unchanged to him]. This produces an observed, but not actual modification of the absolute, actual contraction.

[Of course, if one of the moving clocks is moving at a modest velocity the difference between its at rest dimensions and its actual contracted ones is so small that the observations from that slow-moving clock would be essentially equivalent to from at rest, the very case set out for planet Earth in (3) above.]

In his original paper on relativity Einstein contended that there was no way that an observer experiencing acceleration could distinguish between whether his system was actually accelerating in a region free from gravitation or was actually at rest in a gravitational field. In fact, that contention is incorrect and the distinction can be made by local measurement, as is now known. The distinction occurs because gravitation follows an inverse square law in practice in the real universe and gravitation is inherently radial relative to the gravitating mass.

One could say that Einstein was largely correct but for partially incorrect reasons. The same can be said of the effect of absolutivity on cosmology and space-time physics. The results obtained by traditional 20$^{th}$ Century physics and the theories leading to them are largely correct. Absolutivity only restores the medium and the prime [but not "preferred", special, nor having different physical laws] frame of reference.

The fact that until recently we could detect no absolute velocity and that even now it is only detectable with special scientific effort does not mean that all motion is relative, it only means that we have not developed the means for ready detection of absolutivity. There have been many other things that were undetectable in the past but that are not so now: germs, distant stars, x-rays, atoms, etc.

The Theory of Relativity has required mind-twisting adjustments to way of thinking, adjustments away from the reasonable and "apparent" to a mass of paradoxes and their proposed resolutions. Absolutivity retains contact with reality both in describing physical reality accurately and by doing so in a fashion much more consistent with reasonableness.

With absolutivity the principle of invariance becomes simple, practical and apparent in addition to being necessary as it always was. There is only one "system", the universe with some parts moving in various ways and some parts at rest and that one system has, of course, one overall set of physical laws throughout. Before absolutivity, invariance was necessary but was crying for an explanation, a cause. One can see no particular reason why invariance should be necessarily automatically true in the universe of the Theory of Relativity. Absolutivity solves the problem by showing the natural inevitability of invariance.

Why does this new medium succeed when all prior attempts to define an "aether" without contradictions failed? The reason is the nature of the electro-magnetic wave medium, as follows.

Electro-magnetic field is cyclically changing electric and magnetic field. It is caused by changing motion of electric charge. The changes are changes in the always-present static field of electric charge. The variations in the static field are relative to its average value, the static field amount in the absence of motion of charge. The magnetic field is a further variation in the static field, a distortion of it due to the effect of charge motion.

Static electric field is normally thought of as just that, static. But, if electro-magnetic waves are merely variations in that field and yet they propagate at the speed of light, then the static electric field must be a propagation of some thing at the speed of light, $c$. Such a propagation model of static electric field is essential. Otherwise communication at speeds in excess of the speed of light could take place by making static field changes. See *Inertial Mass, Its Mechanics - What It Is; How It Operates* [2].

That propagation, the static electric field, is the medium, the "aether", and it is relative to the universe' prime frame of reference, that of the "Big Bang", that of where its source charges originated, where they were before motion carried them elsewhere. That propagation emanates from each charge, originally from the origin of the "Big Bang" and now from wherever each charge is.  It, itself carries the controlling parameters  $\mu$  and  $\varepsilon$.   See *Gravitational Mass, Its Mechanics - What It Is; How It Operates* [3].

It is now time to address the apparent paradox that was left as a question at the beginning of this discussion.  The apparent paradox had two elements.

> First
>> A charge at rest relative to the Earth's surface exhibits to us, who are also at rest relative to the Earth's surface, no magnetic field even though the charge is clearly in motion with the Earth's surface rotating about the planet's axis, revolving about the sun and moving relative to and with the galaxy.
>
> Second
>> A charge in motion in an electric wire [as a current] does exhibit a magnetic field to us, who are [in this problem] moving with the same velocity as the charge even though the charge is at rest relative to us.

Although there are these two elements to the problem, they are one overall problem, an apparent inconsistency in physical laws.  The inconsistency results directly from relativity and resolves when absolutivity is applied.

> Considering first the problem of the wire, absolutivity answers with the solution,
>> "Since the current in the wire is in absolute motion, it exhibits the usual magnetic field regardless of the motion of the observer.  The only effect of the observer's motion is to change his standards of measurement and, therefore, the magnitude of the magnetic field as he measures it."

Relativity responds,

> "No, the explanation is that, although the current of the charge moving relative to the wire is zero relative to the observer moving at the same velocity, the overall wire including the charge is electrically neutral so that the wire moving 'rearward' without the charge [as the observer sees it] is an opposite charged wire moving in the opposite direction and produces the same magnetic field to the observer as he would see if he were at rest relative to the wire and he were observing the charge moving 'forward'.  In other words, a wire moving 'rearward' while its current stands still gives the same magnetic field as the wire standing still and its current moving 'forward'."

Absolutivity then closes the discussion with,

> "If relativity were valid that would be a true and good analysis, but the same problem as that of the wire can be stated for a beam of charged particles in empty space without the wire.  In such a case the magnetic field behavior is the same, the paradox for relativity is the same, but there is no 'wire' to travel 'rearward'.  Thus, only the explanation of absolutivity will resolve the problem."

[This also illustrates the simplicity of absolutivity as compared to the complications of relativity.]

The first part of the paradox, that of the charge at rest on the Earth's surface, is simply a case of magnitudes.  In fact the charge at rest relative to the moving Earth is in absolute motion and does exhibit the expected magnetic field.  However, the field is too small to be noticed.  The magnitude of magnetic field is less than the corresponding electric field magnitude by a factor of $[v^2/c^2]$, the $v$ being the velocity of the charges whose motion, as electric current, produces the magnetic field.  The velocity of Earth [presented earlier above] is less than $10^{-3}$ of the speed of light so that $[v^2/c^2] < 10^{-6}$.

In addition, of course, the Earth is overall electrically neutral and the magnetic field due to its motion in space consists largely of a pair of equal and opposite such fields.

## PHILOSOPHIC COMMENTS ON THE NEW RESULT: "ABSOLUTIVITY" REPLACING RELATIVITY

Science on the large scale, that is science dealing with the fundamentals of reality and the universe, has always had and still has a major effect on the non-scientific - social - general philosophic thinking of that science's society and its leaders.

The beginning of the scientific method and the work of scientists such as Copernicus and Galileo resulted in the new period of "The Age of Reason" and "The Enlightenment" – rationality and empiricism replacing dogma and faith.

The new developments that Newton introduced led directly to the concept of the "clockwork universe" and the strong belief in laws, order and regularity.

And, Einstein's theory of relativity coupled with the 20$^{th}$ Century's attribution of actual uncertainty or indeterminism to all physical objects, an extension far beyond the original valid Heisenberg Uncertainty of measurement due to the act of measuring changing the object measured, resulted in our contemporary outlook of a probabilistic reality with no certainty, everything relative with no firm truths, upon which we can lay some of the responsibility for the horrors and tragedies of the 20$^{th}$ Century.

How is that so ?

---

In general, a statement and its contradiction cannot be simultaneously true. Therefore, there are some absolute truths. Thus there is absolute truth, which is the collective body of absolute truths.

Not all statements are absolute truths. Aside from error, which by definition is not true, there is opinion. For example:

- Some people state their liking for candy; some their dislike. It is a matter of opinion.
- But, the statement "Some candy has properties that appeal to some people" is an absolute truth.

The point of view that the questions, "What is truth ?" and "What is real ?" are meaningless questions without answers is not only incorrect but quite negative and harmful in that it suppresses inquiry and progress that could otherwise take place.

Truth is that which conforms to and describes reality. Reality is that which is, not only matter and energy in their various forms but also: feelings and emotions, ideas and cultures, languages and arts, and so forth.

**Whether we can know, sense, measure, or understand some aspect of reality or not it still, nevertheless, is.**

Its being does not depend on our consent nor our observation nor our understanding of it, nor even our own being. We are not gods.

The problem is not whether there is absolute truth or not -- there is. The problem is finding out, coming to know, what the absolute truth is, what is true and what is not. Just what is the "real" reality.

This problem has beset mankind since the earliest stages of the development of our reasoning. It has resulted in a more or less collective decision to grant equal validity to a number of different versions of the truth in spite of their being mutually contradictory.

Not that individuals, organizations and governments hold the opinion that their own version of the truth is not correct. Rather, they ardently believe in the correctness of their own views. But, their inability to prove their views and their inability to defeat differing or opposing views necessitates their getting along in some fashion with those other views and the multiplicity of contradictory views of reality.

That state of affairs has existed for so many human lifetimes that it has essentially implanted in our collective and individual thinking the incorrect belief that there is no absolute truth, that truth is what we say it is -- especially that truth is what we can <u>enforce</u> it to be.

We have gone from inability to determine the truth to non-belief in its existence and then to belief that truth, and reality, are whatever we choose to believe them to be and can force on our fellows.

The most significant characteristic of the 20th Century, other than its explosion of technology, has been its adoption of the attitude that truth is different for each person and each case, that it is what each individual perceives it to be -- that there is no objective reality, only the subjective reality as perceived by each individual -- that all is relative.

The great damage that such thinking does is the license that it gives. It gives license to create, choose, decide upon one's own "reality" and then act accordingly. Such thinking ultimately gives us war, rapine, holocausts.

But, if there is an absolute reality, objective truth, then, even if we are not able to completely know and understand it, we are subject to it. We are measured and judged by it; we experience the effects and consequences of it whether we agree and approve or not, and we feel compelled to behave accordingly.

*Thus absolute reality and objective truth,*
*which indeed exist,*
*also are desirable and beneficial.*

*They are, in fact, essential to civilized society.*

And, that is the beneficial result of Absolutivity replacing Relativity.

## *References*

**[1]** This paper is based on development in R. Ellman, *The Origin and Its Meaning*, The-Origin Foundation, Inc., http://www.The-Origin.org, 1997, in which there is more extensive development and the collateral issues are developed.

**[2]** R. Ellman, *Inertial Mass, Its Mechanics - What It Is; How It Operates*, Los Alamos National Laboratory Eprint Archive at http://arxiv.org, physics/9910027.

**[3]** R. Ellman, *Gravitational Mass, Its Mechanics - What It Is; How It Operates*, Los Alamos National Laboratory Eprint Archive at http://arxiv.org, physics/9903035.